\documentclass[11pt]{cernrep}
\usepackage{graphicx}
\usepackage{mathbbol}
\usepackage{amsmath,amssymb}
\usepackage{axodraw}

\newcommand{\bra}{\langle}
\newcommand{\ket}{\rangle}

\newcommand{\D}{\mathcal{D}}

\newcommand{\tr}{\mbox{tr}}

\newcommand{\ssvc}[1]{\mbox{\scriptsize\boldmath$#1$}}

\newcommand{\be}{\begin{equation}}
\newcommand{\ee}{\end{equation}}
\newcommand{\bea}{\begin{eqnarray}}
\newcommand{\eea}{\end{eqnarray}}
\newcommand{\bean}{\begin{eqnarray*}}
\newcommand{\eean}{\end{eqnarray*}}
\newcommand{\bml}{\begin{mathletters} \baselineskip 10pt}
\newcommand{\eml}{\baselineskip 12pt \end{mathletters}}

\begin{document}

\title{EFFECTIVE LATTICE ACTIONS FOR FINITE-TEMPERATURE YANG-MILLS THEORY}

\author{Thomas Heinzl$^{1}$, Tobias K{\"a}stner$^{\,2}$ and Andreas Wipf$^{\,2}$}

\institute{$^{1}$ School of Mathematics and Statistics, University
of Plymouth, Drake Circus, PL4 8AA, UK\\
$^{2}$ Theoretisch-Physikalisches Institut, FSU Jena,
Max-Wien-Platz 1, 07743 Jena, Germany}

%

\maketitle

\begin{abstract}
We determine effective lattice actions for the Polyakov loop using
inverse Monte Carlo techniques.
\end{abstract}

\section{Introduction}

As there are different notions of effective actions let us start
right away with the definition we will employ throughout this
presentation: Given an action $S = S[U]$ for some
\textit{`microscopic'} degrees of freedom $U$ we define an action
for \textit{`macroscopic'} degrees of freedom $X \equiv X[U]$ via
the functional integral
\be \label{DEF}
    e^{-S_{\mathrm{eff}}[{X}]} = \int \D \, U \, \delta({X} -
    {X}[U]) \, e^{-S[U]} \; .
\ee
Hence we `integrate out' ${U}$ in favor of ${X}$ which will
guarantee that the actions ${S}$ and ${S_{\mathrm{eff}}}$ have the
same matrix elements for the remaining degrees of freedom ${X}$.
Typical examples of such effective actions are obtained if ${X}$
corresponds to low-energy degrees of freedom, like in chiral
perturbation theory. Alternatively, ${X}$ may represent some order
parameter (field) as is common in Ginzburg-Landau theory, for
instance.

While this is all fairly straightforward in principle one
encounters difficulties in practice: in general, the
${U}$-integration cannot be done analytically. Fortunately, there
is a particularly elegant way out, encoded in the `effective field
theory' program. There, one argues that the effective action
${S_{\mathrm{eff}}[{X}]}$ should have the same symmetries as the
`parent' action ${S[U]}$ which suggests the ansatz
\be \label{S_EFF_ANSATZ}
  S_{\mathrm{eff}}[{X}] = \sum_k {\lambda_k} \, S_k [{X}] \; ,
\ee
representing a systematic expansion in symmetric operators ${S_k}$
of increasing mass dimension (multiplied by inverse powers of the
cutoff). The parameters ${\lambda_k}$ are then fixed via
`phenomenology'.

Alternatively, one may try to `do the ${U}$-integration
numerically' on a lattice. However, on a lattice one can only
calculate \textit{expectation values}. So the question arises: How
can one find effective actions from the latter? The answer is
given in the next section.

\section{Inverse Monte Carlo}

Inverse Monte Carlo (IMC) avoids the functional integration in
(\ref{DEF}) by a detour consisting of three basic steps: first,
generate configurations $U^{(i)}$, $i=1, \ldots, N$, via standard
MC procedures. Second,  calculate the configurations ${\big\{
{X}^{(i)} \big[ U^{(i)} \big] \big\}}$ and compute the expectation
values $\bra O[{X}] \ket_S = N^{-1} \sum_{i=1}^N O \big[{X}^{(i)}
\big]$. Note that the $X$ variables are distributed according to
$\exp(-S)$, i.e.\ the parent action. Third, in the IMC step
proper, determine the effective couplings ${\lambda_k}$ via
Schwinger-Dyson equations (SDEs) \cite{falcioni:86,dittmann:2002}.
The latter step requires some explanations.

Generically, the target space (where the $X$ variables `live')
will have some isometry leaving the metric invariant,
${{\hat{\mathcal{L}}_X} \, g^{ab}(X) = 0}$, $\hat{\mathcal{L}}$
denoting the relevant Lie derivative. This entails an invariance
of the functional measure, ${\D {X} \equiv \prod d{X} \,
\sqrt{g({X})}}$, leading to path integral identities that can be
cast into the SDEs
\be
  \sum_k \bra F {\hat{\mathcal{L}}_X} S_k \ket_S {\lambda_k} = \bra
  {\hat{\mathcal{L}}_X} F \ket_S \; ,
\ee
where the ansatz (\ref{S_EFF_ANSATZ}) has been utilised. The
former constitute an (overdetermined) linear system for the
${\lambda_k}$ which can be solved numerically. The arbitrary
function ${F[{X}]}$, if properly chosen, can be used to fine-tune
the associated numerics. An expensive (!) check of the solution
can (and should) be done by testing the equality of the matrix
elements calculated with both the parent and daughter actions,
$\bra O \ket_{S_{\mathrm{eff}}} \stackrel{?}{=} \bra O \ket_S$.

\section{Polyakov-Loop Dynamics}

In \cite{dittmann:2003,heinzl:2005} we have considered some
particular examples of effective actions for the Polyakov loop
based on suggestions in
\cite{svetitsky:86,billo:1996,pisarski:00}.

\subsection{Generalities}

At the risk of boring the experts we briefly review the physics of
the Polyakov loop. This quantity is all-important for finite
temperature Yang-Mills theory as it constitutes a gauge invariant
order parameter for the confinement-deconfinement phase
transition. The Polyakov loop is a traced holonomy (or Polyakov
line),
\be
  {{L}_{\ssvc{x}} [U] \equiv \frac{1}{N_C} \tr_F \,
  {\mathfrak{P}}_{\ssvc{x}} [U]} \; , \quad
  {\mathfrak{P}}_{\ssvc{x}}[U]
  \equiv \prod_{t=1}^{N_t} U_{t,\ssvc{x};0} \; .
\ee
Introducing the Wilson coupling by $\beta = 2 N_C / g^2$ the
theory assumes its confinement phase for $\beta < \beta_c$ with
${\bra {L}\ket = 0 }$ and its deconfinement phase for ${\beta
> \beta_c}$ with nonvanishing expectation value, ${\bra {L}\ket \ne 0 }$.
The latter phase is characterised by a spontaneously broken centre
symmetry generated by transformations under which $L$ transforms
nontrivially, $L_{\ssvc{x}} \to z \, L_{\ssvc{x}}$, $z \in
\mathbb{Z}(N_C)$.

The critical behaviour is characterised by the Svetitsky-Yaffe
conjecture \cite{svetitsky:82,yaffe:82} which states that the
effective theory describing finite-${T}$ ${SU(N_C)}$ Yang-Mills
theory in ${d+1}$ dimensions is a ${\mathbb{Z}(N_C)}$ spin model
in ${d}$ dimensions with {short-range} interactions. For ${SU(2)}$
this is well established on the lattice via comparison of critical
exponents \cite{fortunato:01,forcrand:01}.

For gauge group $SU(2)$, which we consider henceforth, $L$ is a
real number between $-1$ and 1. This target space is somewhat
nonstandard so that its isometry is not obvious. We therefore
generalise to the group-valued variable, ${\mathfrak{P}} \equiv
P^\mu \sigma_\mu \in SU(2)$. As $SU(2) \cong S^3$ its isometry is
clearly an ${O(4)}$ symmetry generated by `angular momenta'
${M^{\mu\nu}}$. Gauge invariance invariance then implies that the
effective action can only depend on the zeroth component of
$\mathfrak{P}$, ${S_{\mathrm{eff}}[{\mathfrak{P}}] \equiv
S_{\mathrm{eff}}[{P^0}] \equiv S_{\mathrm{eff}}[{L}]}$.
Restricting our identities to $P^0 = L$ results in the following
SDEs for the ansatz (\ref{S_EFF_ANSATZ}),
\be
  \sum_k \bra (1 - {L}_{\ssvc{x}}^2) G S_{k,\ssvc{x}} \ket
  {\lambda_{k}} = \bra  (1 - {L}_{\ssvc{x}}^2) G_{, \ssvc{x}} -
  3 {L}_{\ssvc{x}} G \ket \; ,
\ee
where ${ G_{, \ssvc{x}} \equiv \partial G/\partial
{L}_{\ssvc{x}}}$ etc. Again, the functional $G$ is useful for
fine-tuning. An optimal choice is to use the derivatives of the
effective action, ${G \in \{ S_{l, \ssvc{y}} \}}$, which are the
operators appearing in the equation of motion for $L$. In this
case the SDEs become relations between two-point functions of the
Polyakov loop $L$. Again, they represent an exact, overdetermined
linear system for the couplings ${\lambda_{k}}$. We still have not
decided upon the concrete form of the effective action. This is
somewhat of a problem, as the former is only mildly constrained by
centre symmetry. The additional fact that $L$ is dimensionless
allows for a plethora of operators $S_k$. Certainly, a further
guiding principle is needed.

\subsection{Effective Action from Character Expansion}

Recall that any (gauge invariant) function on ${SU(N_C)}$ can be
expanded in terms of an adapted `Fourier' basis, the group
characters for representation $R$, $\chi_R [U] \equiv \tr_R \, U$,
with $U \in SU(N_C)$.
For ${SU(2)}$ we use the notation ${\chi_p}$ with  ${p = 2j}$ and
${j = \frac{1}{2}, 1 , \frac{3}{2}, \ldots,}$ denoting `color
spin'. The lowest characters are the polynomials ${\chi_1} =
2{L}$, $\chi_2 = 4 {L}^2 -1$, $\chi_3 = 8 {L}^3 - 4 {L}$,
$\ldots$. If we impose ${\mathbb{Z}(2)}$ symmetry  and
reducibility of product representations we obtain the following
systematic expansion for the effective action (see also
\cite{billo:1996,dumitru:03}),
\be
  S_{\mathrm{eff}}[{\chi}] = \sum_{\bra \ssvc{x} \ssvc{y}\ket}
  \sum_{\scriptstyle pq \atop \scriptstyle {p - q \, \mathrm{even}}} \lambda_{pq}
  \, {\chi_p}
  [L_{\ssvc{x}}]  \, {\chi_q}  [L_{\ssvc{y}}]
  + \sum_{\ssvc{x}} \sum_{{p \, \mathrm{even}}}
  \lambda_{p0} \,
  {\chi_p} [L_{\ssvc{x}}]  \; . \label{S_EFF_CHAR}
\ee
This action is a ${\mathbb{Z}(2)}$-symmetric sum over higher and
higher ${SU(2)}$ representations including both nearest-neighbour
(NN) hopping and potential terms. Before determining this action
numerically it is worthwhile to study its consequences in
mean-field approximation (MFA).

\subsection{Mean-Field Approximation}

\begin{figure}
  \includegraphics[scale=0.6]{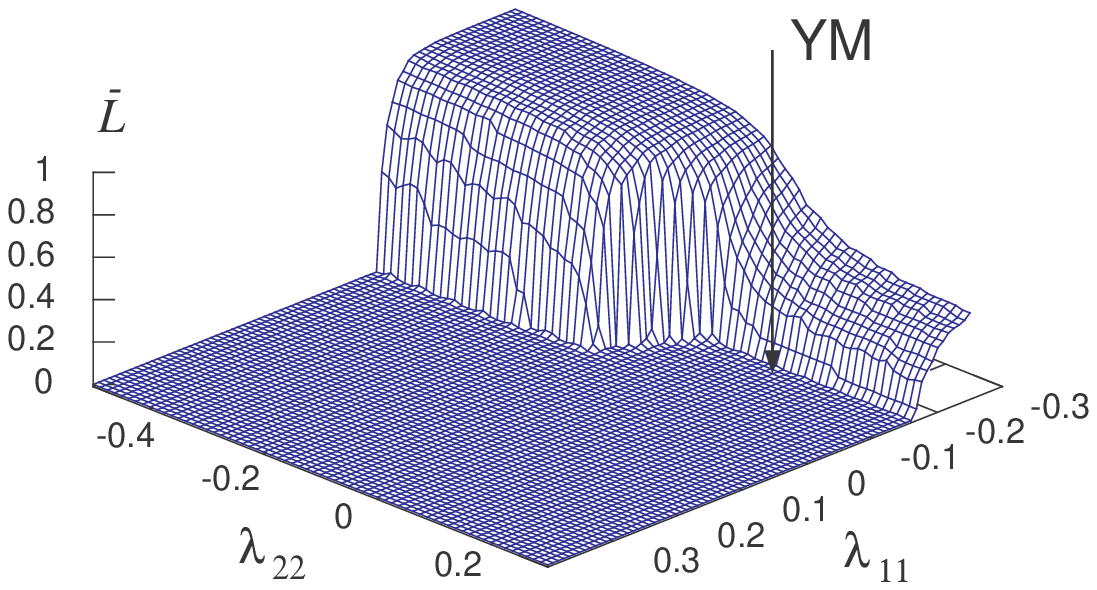}
  \hfill
  \includegraphics[scale=0.6]{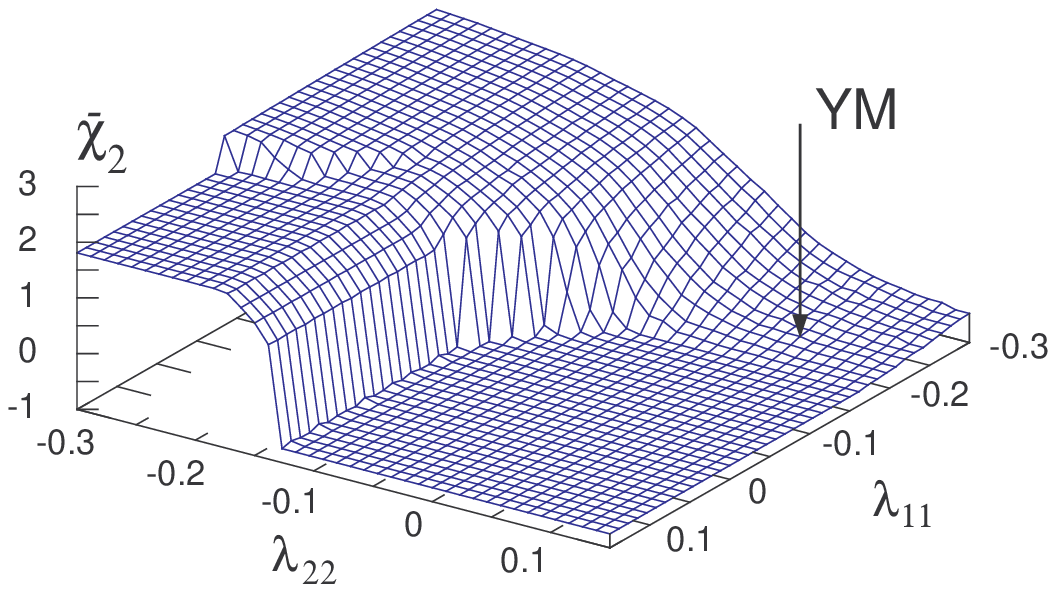}
  \caption{\label{FIG:CHIMEAN}The MF predictions for the VEVs of the
  Polyakov loop  $\bar{L} = \bar{\chi}_1 /2$ (left) and the
  adjoint character $\bar{\chi}_2$ (right) as a function of the couplings
  $\lambda_{11}$ and $\lambda_{22}$.}
\end{figure}

If we restrict ourselves to the fundamental and adjoint
representations (${p=1,2}$) and to hopping terms only we end up
with the effective action (\ref{S_EFF_CHAR}) truncated such that
it contains only couplings $\lambda_{11}$ and $\lambda_{22}$. This
action will be denoted $T_2$ and implies a MF potential
\[
  V_{\mathrm{{MF}}}({\lambda}, {\chi}) \equiv -d \sum_{p=1,2} {\lambda_{pp}}
  {\chi_p}^2 - \log z({\lambda}, {\chi})
\]
where $z$ is the single-site partition function. The two MF (or
gap) equations are $\partial V_{\mathrm{{MF}}} / \partial {\chi_p}
= 0$, $p = 1,2$, the solution of which yields the vacuum
expectation values (VEVs) ${\bar{\chi}_p}$ of the characters. The
latter are displayed in Fig.~\ref{FIG:CHIMEAN}. The behaviour of
$\bar{L}$ as a function of the coupling $\lambda_{11}$ implies the
typical second-order phase transition (for $\lambda_{22}$
sufficiently large). Interestingly, the adjoint character
$\bar{\chi}_2$ displays discontinuous behaviour as a function of
$\lambda_{22}$ (for $\lambda_{11}$ sufficiently large). This
behaviour, however, is located off the `physical region' marked by
the arrows. These point towards the couplings obtained by
numerically matching to the Yang-Mills action via IMC. This is our
next topic.

\subsection{Inverse Monte Carlo}

\begin{figure}
  \includegraphics[scale=0.7]{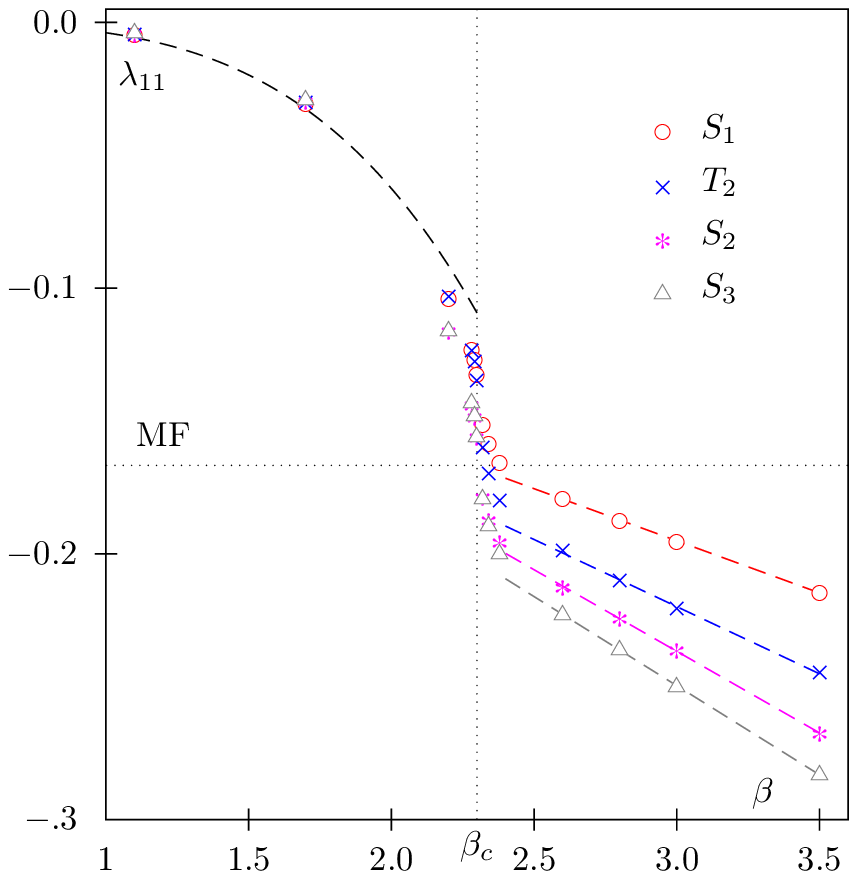}
  \hfill
  \includegraphics[scale=0.7]{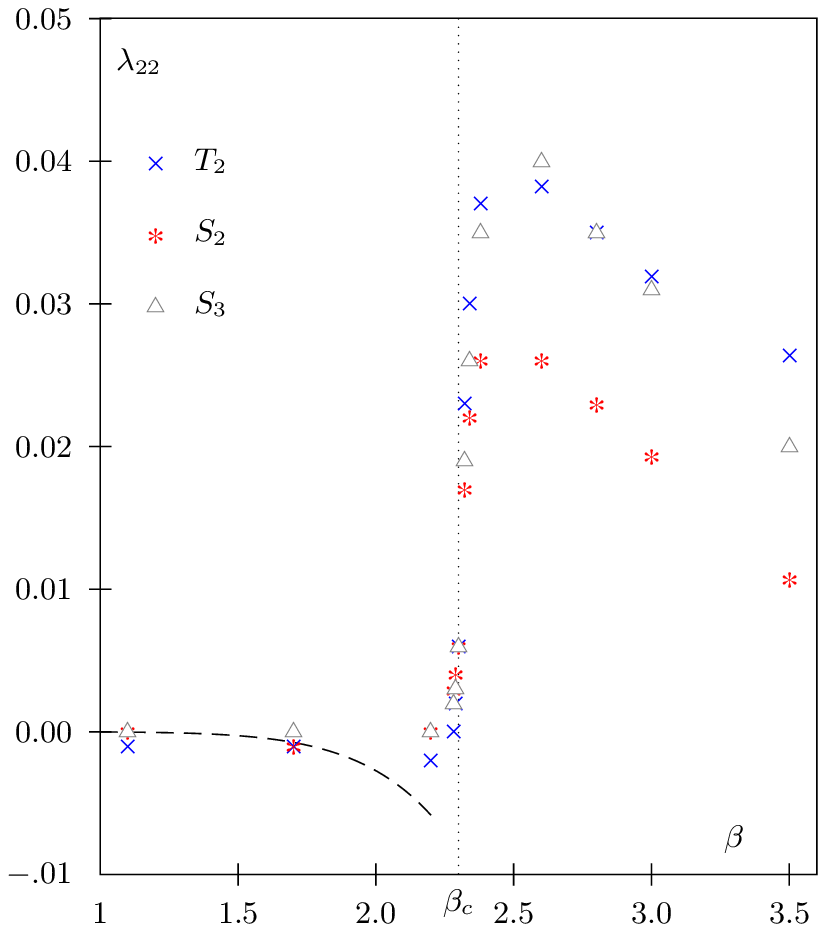}
  \caption{\label{FIG:EFF_COUP}Effective couplings determined via IMC.
  Left: fundamental coupling ${\lambda_{11}}$,
  Right: adjoint coupling ${\lambda_{22}}$. Different symbols correspond to different
  truncated actions, $S_1, \ldots , S_3$, see text. Dashed lines
  for small $\beta$ are strong-coupling results.}
\end{figure}

We have simulated $SU(2)$ Yang-Mills theory on a  $4 \times 20^3$
lattice for Wilson coupling $\beta$ ranging between $1.2$ and
$3.5$. The critical coupling is $\beta_c = 2.30$. The associated
ensembles have been matched to effective actions $S_p$ with
representations $p \le 3$, hence including up to five character
terms and, accordingly, five effective couplings. Some results are
presented in Fig.~\ref{FIG:EFF_COUP}.

%
%

We see that both fundamental and adjoint couplings ($\lambda_{11}$
and $\lambda_{22}$) jump near the critical $\beta$. The effect of
including potential terms and higher representations ($S_2$,
$S_3$) is rather mild and visible mainly in the broken phase
($\beta > \beta_c$). The linear behaviour of $\lambda_{11}$ there
is predicted by perturbation theory \cite{ogilvie:84}. Finally,
the MF critical coupling for an Ising type effective action $S_1$
($\lambda_{11}$ only) is $\lambda_{11c} = - 1/2d = -0.17$ is off
by only a few percent.

\begin{figure}[!h]
  \includegraphics[scale=0.7]{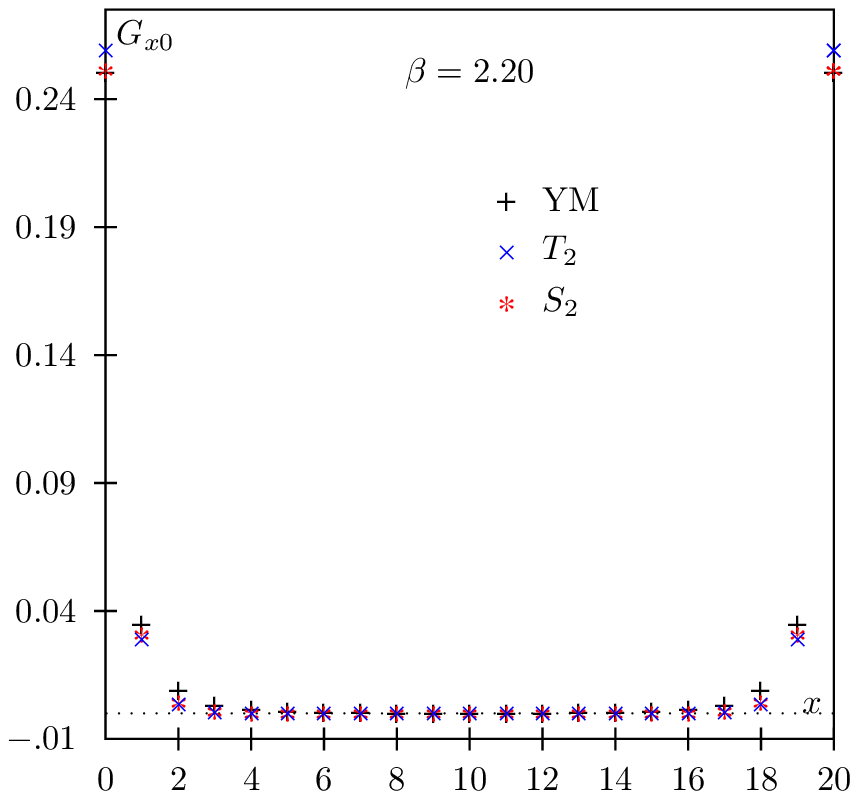}
  \includegraphics[scale=0.7]{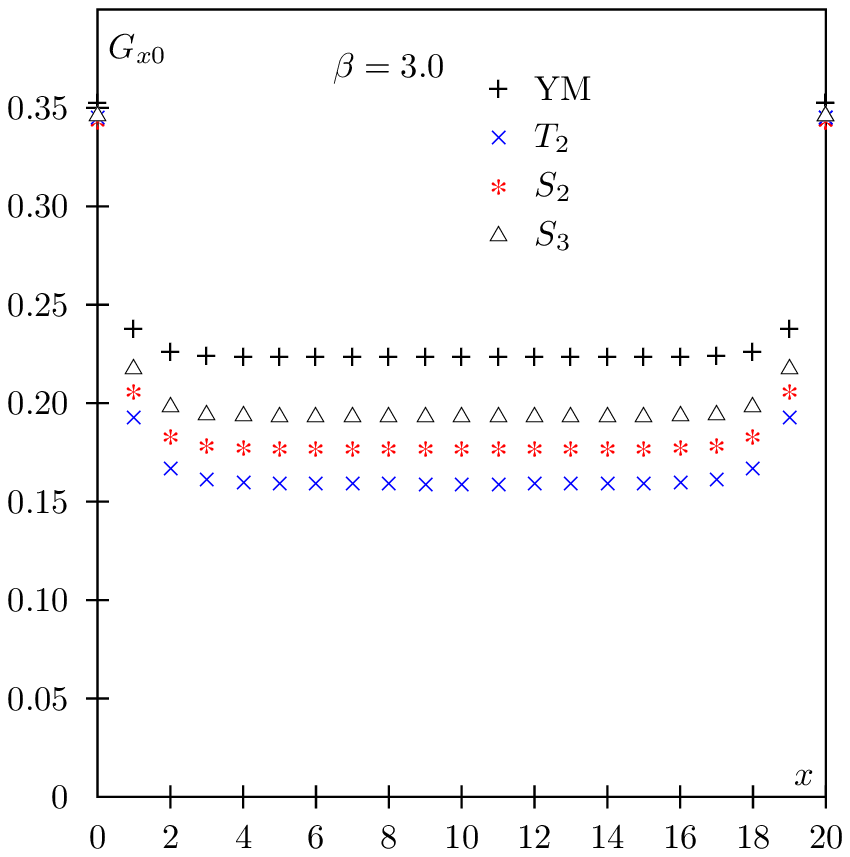}
\caption{\label{FIG:TWOPOINT}Two-point functions for $\beta = 2.2$
(left) and $\beta= 3.0$ (right).}
\end{figure}

To check the quality of our effective actions we have simulated
them using the IMC values for the couplings and compared simple
matrix elements namely the two-point functions (see
Fig.~\ref{FIG:TWOPOINT}).

We note that in the symmetric phase ($\beta = 2.2$) the data for
different truncations lie on top of each other and reproduce the
Yang-Mills values. In the broken phase ($\beta = 3.0$), however,
the situation is different. Including higher representations leads
to improvement but in particular the Yang-Mills plateau value
(hence $\bra L \ket$) is not reproduced.

\begin{figure}
  \includegraphics[scale=0.55]{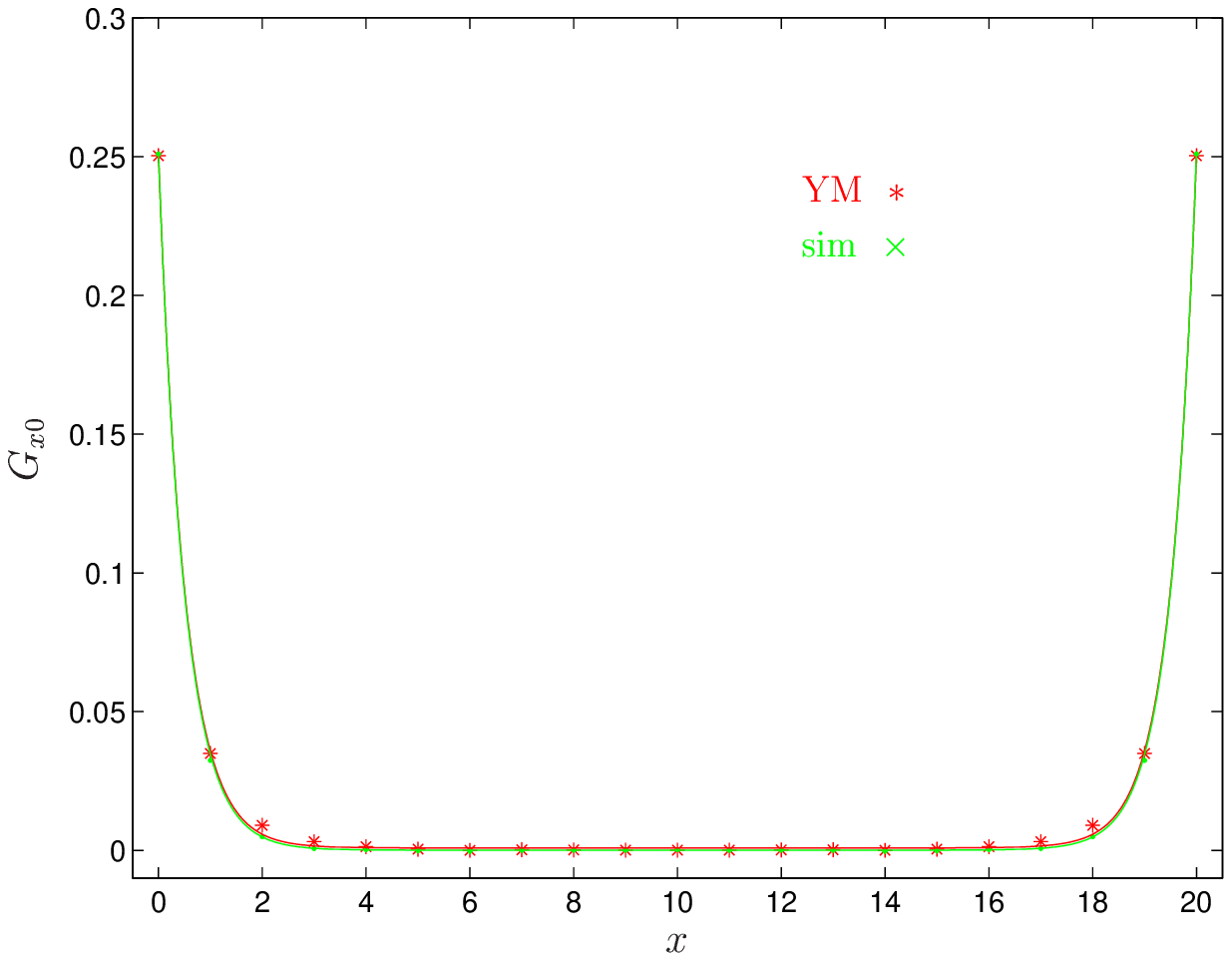}
  \includegraphics[scale=0.55]{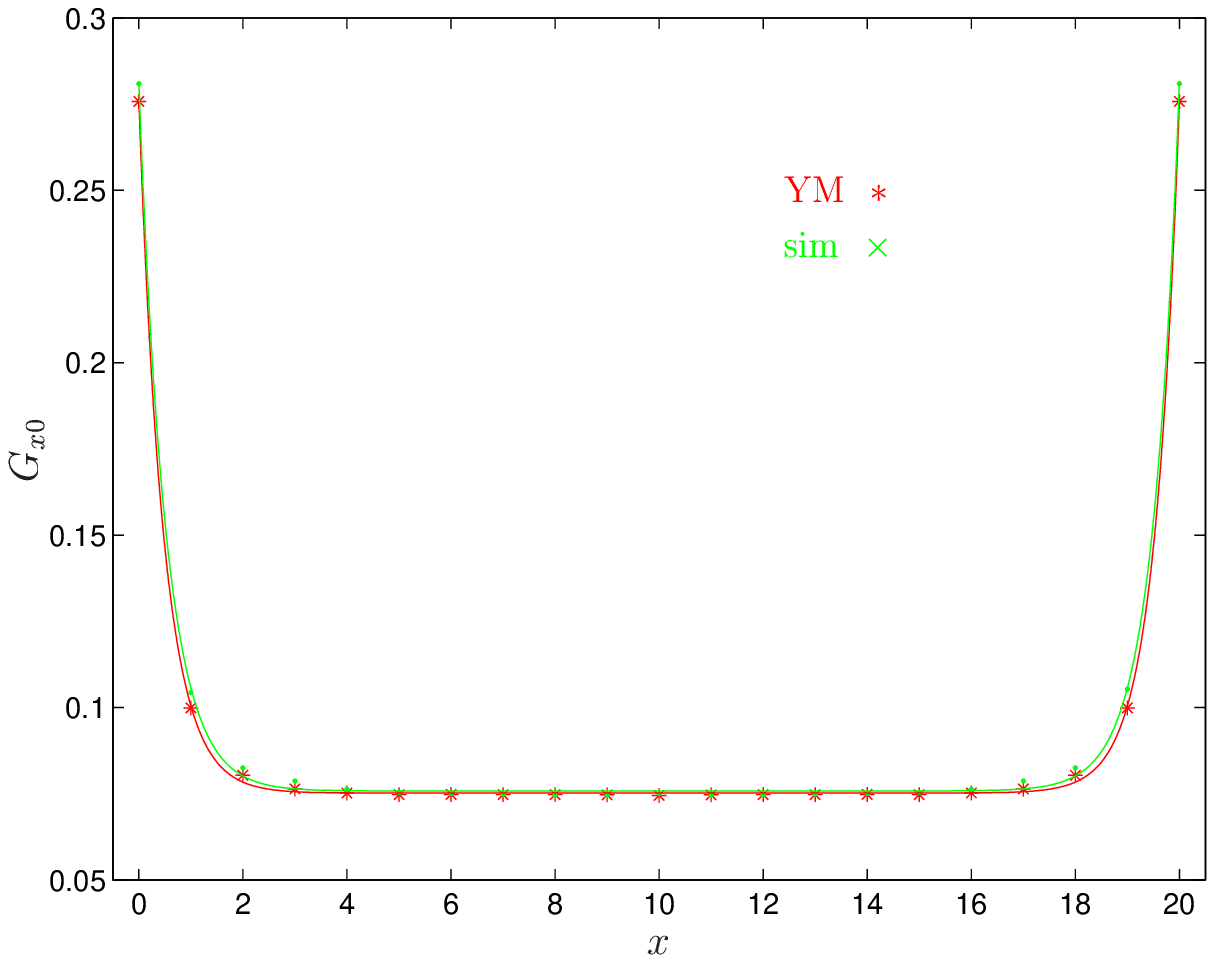}
  \caption{\label{FIG:TWOPOINT14}Two-point functions for $\beta =
  2.2$ (left) and $\beta= 2.4$ (right) using 14 NN and NNN operators
  (``sim").}
\end{figure}

To remedy this fault we have performed a brute-force calculation
by also including next-to-nearest neighbour (NNN) interactions and
a total of 14 different operators \cite{dittmann:2003}. We find
that generically the effective couplings decrease with the
representation label $p$ (as above) and the distance of the sites
connected by the operators. A comparison of the two-point
functions now yields a near-perfect match (see
Fig.~\ref{FIG:TWOPOINT14}).

%

\section{Summary and Outlook}

We have seen that inverse Monte Carlo based on Schwinger-Dyson
equations can be a powerful method to numerically determine
effective actions. Applying it to the Polyakov loop dynamics in
${SU(2)}$  yields ${\mathbb{Z}(2)}$-symmetric effective actions
${S_{\mathrm{eff}}[{L}]}$ describing the confinement-deconfinement
phase transition in quite some detail. The matching to Yang-Mills
is reasonable if NN character interactions are used and becomes
near-perfect if one includes NNN interactions. Obvious extensions
of this work will be to go to $SU(3)$ and to analyse explicit
symmetry breaking terms (mimicking `fermions'). A study of the
continuum limit will necessitate to renormalise the lattice
Polyakov loop.

\section*{Acknowledgments}
It is a pleasure to thank the organisers of XQCD, G.~Aarts,
C.~Allton, S.~Dalley, S.~Hands and S.~Kim for the great job they
have done.

\enlargethispage{\baselineskip}

\end{document}